\documentclass[12pt]{iopart}
\usepackage{iopams}

\usepackage[dvips]{hyperref}
\usepackage{xcolor}

\usepackage{euscript,amssymb,graphicx,amsfonts,latexsym,epsfig}
\usepackage{subfigure}

\usepackage{graphicx}
\newcommand{\nn}{\nonumber}
\newcommand{\eqref}[1]{((\ref{#1}))}
\newcommand{\eqr}[1]{Eq.~(\ref{#1})}
\newcommand{\m}{\mathbf}
\newcommand{\f}{\frac}

\begin{document}

\title{Stable sound wave generation in weakly ionized air medium}

\author{Maxim Chizhov\dag}
\ead{max-lopez@ukr.net}

\author{Maxim Eingorn\dag\ddag \S}
\ead{maxim.eingorn@gmail.com}

\author{Vladimir Kulinskii\dag}
\ead{kulinskij@onu.edu.ua}

\

\address{\dag Faculty of Physics, Odessa National University,\\ Dvoryanskaya st. 2, Odessa 65026, Ukraine}

\

\address{\ddag Department of Theoretical and Experimental Nuclear Physics,\\ Odessa National Polytechnic University,\\ Shevchenko av. 1, Odessa 65044, Ukraine}

\

\address{\S Physics Department, North Carolina Central University,\\ Fayetteville st. 1801, Durham, North Carolina 27707, USA}

\

\begin{abstract}
We consider the generation of sound waves in the air medium  between electrodes at the voltages near electrical breakdown in the presence of the time dependent
constituent of the electric field. Within the standard multicomponent hydrodynamic model of the weakly ionized gas it is shown that the generation of sound is
possible due to instantaneous character of the ionization equilibrium. The influence of the electronegative ions on the sound intensity is also discussed.
\end{abstract}

\maketitle

\section{Introduction}\setcounter{equation}{0}

The acoustic devices based on the plasma and the corona discharge have important advantage of direct transformation of the electric signal into  sound. This
opens up possibilities for designing the ``membrane-free`` acoustic systems
\cite{plasma_coldsound_jacsocam1973,plasma_acoustcorona_applacoust1981,plasma_acoustloudspeakers_jphysd1987} in contrast to the standard loud-speakers
\cite{book_loudspeakers}. The main principle of operation of the electroacoustic transducers is the transformation of the electric energy into oscillation of
the air density (pressure). Such transformation can be either direct or indirect. The direct transformation is when the electric energy transforms into the
motion of macroscopic volumes of the media producing the change of the pressure. The indirect one is connected with the change of the internal energy, i.e. the
heat dissipation by the Joule losses and the corresponding thermodynamic process of changing the pressure (density). With respect to these mechanisms the
transducers are divided into two classes: the ``cold`` and the ``hot`` ones \cite{plasma_acoustloudspeaker_nrs1982,plasma_acoustloudspeaker2_nrs1982}. The
heating mechanism of the sound generation is typical for the high frequency corona type and plasma  sources
\cite{plasma_acoustcorona_applacoust1981,plasma_loudspeaker_japp2001,plasma_loudspeaker_jacsocam2007}. The force mechanism of the sound generation is
characteristic for the low frequency ($<10\,kHz$) corona sources
\cite{plasma_coldsound_jacsocam1973,plasma_loudspeaker_japp2001,plasma_loudspeaker_jacsocam2007}. The simplified electric circuit model was developed to
describe the main characteristics of the transducers \cite{plasma_coldsound_jacsocam1973,plasma_loudspeakerdisch_jacsocam1987}. Basically monopolar and dipolar
directivities are associated to the ionization and drift regions, respectively.  The physics behind the above-mentioned devices is known and the main
predictions concerning the directivity pattern match very well with the experimental results. In particular, the effective electrical circuit models were
proposed \cite{plasma_coldsound_jacsocam1973,plasma_loudspeaker_japp2001}. They give a clear physical picture on the phenomenological level and describe the
main characteristics rather good using the parameters of the effective electric circuit. Nevertheless, there is the need for the more detailed description of
the sound generation based on the more detailed consideration. In particular, the phenomenological approach fails to answer the questions about the ozone
generation, the dependence of the characteristics on the geometry and the chemical composition.

The possibility of the monopolar source for the cold corona type transducers which would be indistinguishable from the heating related field has been noted in
\cite{plasma_loudspeaker_japp2001}. The glow discharge SLT-Ionophone designed by F.~Fransson
\cite{plasma_acoustionophone_qpsr1971,plasma_acoustionophone_jaas1975} seems to display such possibility. Besides, the fact that the discharge is insensitive
to the magnetic field signifies that the ion wind effect does not matter for the sound generation. Unfortunately, the specific physical mechanism behind this
device was not given by the authors of \cite{plasma_acoustionophone_qpsr1971,plasma_acoustionophone_jaas1975}. Obviously, the ``corona wind`` field source is
incompatible with what is known for this dc glow transducer \cite{plasma_loudspeakerdisch_jacsocam1987}.

In this paper the theoretical scheme of calculation of main characteristics of the sound wave excitation process in the gap between electrodes in the presence
of the alternating electric field constituent is proposed. We use the standard three-component hydrodynamic model of the weakly ionized gas, demonstrating that
the generation of the bulk charge is the monopole sound source which is generally present in cold-plasma transducers. Therefore we identify the physical
process of the monopole sound generation in the force driving corona or glow discharge. In particular, this gives the explanation of the action of
SLT-Ionophone \cite{plasma_acoustionophone_jaas1975}. We also propose our own transducer which is based on this effect.

The paper is structured in the following way. In Section 2 we derive a set of equations describing the electroacoustic system under interest, including
electrodynamic and kinetic equations. Next Sections 3 and 4 are devoted to the experimental setup as well as characteristic values and some simple estimates of
main quantities appearing in the problem, while in Section 5 we discuss the volt-ampere characteristic and autoelectronic emission in a concrete device. Our
main results are summarized in Conclusion. Finally, in Appendix we turn to the simplest possible one-dimensional case and obtain a numerical solution of the
problem.

\section{Set of equations for electroacoustic system}\label{sec_mathstatment}\setcounter{equation}{0}
The description of the system of interest is based on the system of equations which incorporate all main physical processes. They describe the distribution of the
electric field and charges. This set of equations is well known and widely used in modeling of such devices
\cite{plasma_acoustenrgconvers_jphysd1986,plasma_acoustdisch_compmodel2009,plasma_acoustdischwirewire_compmodel2009}.

Because of weak ionization of the air medium, we shall calculate different quantities, characterizing the electroacoustic system, within the limits
of quasi-stationary approximation and the corresponding electrodynamic problem. For the medium with the conductivity $\sigma$ we have the following
system of equations:
\begin{eqnarray}\label{eq_electric}
&{}&\triangle\varphi= -\frac{4\pi}{\varepsilon}\rho\,, \nonumber \\
&{}&\frac{\partial\rho}{\partial t}+\nabla{\bf j}=0\,, \\
&{}&{\bf j}={\bf j}_e+{\bf j}_{-}+{\bf j}_{+}= \sigma{\bf E}\,,\nonumber
\end{eqnarray}
where $\triangle$ denotes the Laplace operator; $\varphi$ and ${\bf E}=-\nabla\varphi$ are the potential and the strength of the electric field
respectively; $\varepsilon$ denotes the relative permittivity; the current density ${\bf j}$ consists of three terms ${\bf j}_e$, ${\bf j}_{-}$ and
${\bf j}_{+}$, corresponding to electrons, negative ions and positive ions respectively; finally, the charge density
\begin{equation}\label{eq_rho}
\rho=e(n_{+}-n_{-}-n_e)\,,
\end{equation}
where $e$ denotes the elementary charge; $n_{+}$, $n_{-}$ and $n_e$ are densities of positive ions, negative ions and electrons respectively. Here
and in what follows we use the CGS system of units. In particular, $e\approx4.8\times10^{-10}\, un.\, CGS$.

The above-mentioned quasi-stationary approximation means that in the zeroth order the strength ${\bf E}={\bf E}_0$, where the stationary constituent
${\bf E}_0$ does not depend on time $t$, while in the first order the strength can be presented in the form
\begin{equation}\label{eq_e0e1}
  {\bf E}={\bf E}_0+\delta {\bf E}(t)\, ,
\end{equation}
where $\delta{\bf E}(t)$ is the alternating constituent (hereinafter also called a "signal") and it is assumed that $\delta{\bf E}(t)\ll {\bf E}_0$.

In such a problem setting the main quantity, which is determined by kinetics of collisions, is the effective conductivity $\sigma$. It is clear that in view of
small mobility of ions the current is caused mainly by the electronic component. For the same reason the ionic component makes the main contribution to the
sound pressure. At that the substantial factor is the generation of electronegative ions of oxygen $O_2^{-}$ in consequence of attachment of electrons, that
must be taken into account in balance kinetic equations and corresponding cross-sections of reactions of birth and annihilation of particles. This part of the
problem is the standard problem of kinetics and supposes consideration of processes of loss and birth of electrons, colliding with molecules and ions
\cite{plasma_electronkinetics_ufn2002en}.

To describe the sound generation, one should supplement the electrodynamic Eqs. \eqref{eq_electric} with the equation of motion of the gas (air), taking into
account the electric force, acting on charges:
\begin{equation}\label{eq_motion}
\frac{\partial{\bf v}}{\partial t}+({\bf v}\nabla){\bf v}=-\frac{\nabla P}{\wp}+\frac{\rho}{\wp}{\bf E}+\ldots\, ,
\end{equation}
where $\wp$ is the mass density; $\ldots$ denotes viscosity terms, which can be neglected under consideration of propagation of sound of typical wave lengths
$1\div 10\, cm$. Since the characteristic frequencies are less than $10\,kHz$ we neglect the heating mechanism of the sound generation which is typical for the
plasma arc and corona-type devices of high-frequency \cite{plasma_acoustcorona_applacoust1981,plasma_acoustloudspeakers_jphysd1987}. Only the so called force
source of the sound is taken into account.

Linearizing \eqr{eq_motion} in a standard text-book manner (see, for example, \cite{book_ll6_en}), we obtain:
\begin{equation}\label{eq_motion1}
\frac{\partial{\bf v}}{\partial t}=-\frac{\nabla P}{\wp_0}+\frac{\rho}{\wp}{\bf E}=-\frac{\nabla P}{\wp_0}+\frac{1}{\mu\wp}{\bf j}\,.
\end{equation}
Here we used the definition of mobility $\mu$: ${\bf j}=\mu\rho{\bf E}$, $\sigma=\mu\rho$.

Let us note that $\rho$, ${\bf j}$, $\sigma$,
$n_{\pm}$, $n_e$ and ${\bf v}$ (as well as $\triangle\varphi$, $\nabla{\bf E}$ and $\nabla P$) are the quantities of the same (first) order of
smallness. Similarly the mass density $\wp$ can be decomposed in the following way:
\begin{equation}\label{wp} \wp=\wp_0+\delta\wp\, , \end{equation}
where $\wp_0$ and $\delta\wp$ are quantities of zeroth and first orders of smallness respectively. Linearizing the continuity equation
\begin{equation}\label{eq_cont}
\frac{\partial\delta\wp}{\partial t}+\wp_0(\nabla{\bf v})=0\,,
\end{equation}
after some straightforward computations from \eqr{eq_motion1} we get:
\begin{equation}\label{eq_sound_main}
\frac{1}{c^2}\frac{\partial^2 P}{\partial t^2}-\triangle P=\frac{1}{\mu}\frac{\partial\rho}{\partial t}+\frac{1}{\mu\,\wp_0}(\m{j}_{0}\cdot\nabla\wp) +
\frac{1}{\mu^2}\, \frac{\partial \mu}{\partial E}\,(\m{j}_{0}\cdot\nabla E) ,
\end{equation}
where $c$ is the speed of sound and $\m{j}_{0}$ is the current value in the stationary state. We take into account that the mobility depends on the strength of
the electric field mainly because of the electronic component. Indeed, the electronic mobility $\mu_e$ depends essentially on $E$ in the interval $1<E<5$ CGS
while the ionic mobilities can be considered as constants (see \cite{book_reiserfizikagazrazrada_en}). Obtained Eq. \eqref{eq_sound_main} clearly demonstrates
that there are three sources of sound of different physical nature. \eqr{eq_sound_main} is the standard equation of acoustic oscillations in the presence
of the external source (representing oscillations of the bulk charge, caused by the presence of the alternating constituent of the electric field). Note that
the terms in the right-hand side of \eqr{eq_sound_main} differ by their symmetry. The first one represents an isotropic source ($l=0$ harmonics), while
two others obviously represent the oscillation on the ion wind background and indeed give raise to the common dipolar contribution ($l=1$ harmonics) into the
acoustic directional pattern \cite{plasma_coldsound_jacsocam1973,plasma_loudspeaker_japp2001,plasma_loudspeaker_jacsocam2007}. The relative magnitudes of these
contributions depend on geometry of the device. Thus, it is possible that the sound is produced mainly due to the oscillation of the bulk charge (the first
term in the right-hand side of \eqr{eq_sound_main}). Yet it is explicitly monopolar in contrast to the statement of
\cite{plasma_loudspeaker_jacsocam2007} that all force sources are dipolar since they are connected with the electric wind. From the physical point of view this
term is caused by the low inertia of the ionization equilibrium. This effect becomes more pronounced in the electronegative media where the formation of large
amount of anions becomes possible. Obviously, the first term exists even if the electric wind is negligible. To our knowledge, all ``cold`` corona sound
sources of point-to-plane geometry considered in the literature are of dipole nature
\cite{plasma_acoustcorona_JAES1957,plasma_coldsound_jacsocam1973,plasma_loudspeaker_japp2001,plasma_loudspeaker_jacsocam2007}. Note that in ``hot plasma``
loudspeakers the monopolar source is due to the thermal heating mechanism \cite{plasma_loudspeaker_jacsocam2007}. Thus, our analysis shows that the monopolar
field source term in generally present in ``cold plasma`` loudspeakers.

Obviously,  \eqr{eq_electric} and \eqr{eq_sound1} should be supplemented with the equations of chemical kinetics, reflecting processes of
generation and loss of charges. Under the ambient conditions and the electric field strengths of the order of the breakdown strengths one can neglect the
processes of diffusion transfer and take into account only collisional contributions. Thus, the system of equations of collisional balance in the "minimal"\
model takes the following form:
\begin{eqnarray} \label{eq_kinetic}
\frac{\partial n_{+}}{\partial t}+\nabla(n_{+}\mu_{+}{\bf E})&=&\nu_{+,e}^{(i)}n_e-\nu_{+,e}^{(r)}n_{+}n_{e}\, ,\nonumber\\
\frac{\partial n_{-}}{\partial t}-\nabla(n_{-}\mu_{-}{\bf E})&=&\nu_{0,e}^{(a)}n_e\, ,\\
\frac{\partial n_{e}}{\partial t}-\nabla(n_{e}\mu_{e}{\bf E})&=&\left(\nu_{+,e}^{(i)}-\nu_{0,e}^{(a)}\right)n_e-\nu_{+,e}^{(r)}n_{+}n_{e}\, \nonumber ,
\end{eqnarray}
where we have taken into account that ${\bf j}_i=q_in_i{\bf v}_i$ and ${\bf v}_i=\pm\,\mu_i{\bf E}$ ($i=\pm,e$); $\nu_{i,j}^{(k)}$ are rates of ionization
($k=i$), recombination ($k=r$) and attachment ($k=a$). In general, mobilities $\mu_i$ depend on $E=|{\bf E}|$. The balance Eqs. (\ref{eq_kinetic}) agree with
the charge conservation law and in our approximation $n_{+}$ and $n_{e}$ are quantities of the first order of smallness, therefore the term
$-\nu_{+,e}^{(r)}n_{+}n_{e}$ in right-hand sides is of the second order of smallness and, consequently, can be neglected. Eqs.~(\ref{eq_electric}),
(\ref{eq_kinetic}) describe the sound generation process under interest. This system should be supplemented with the boundary conditions on surfaces of
electrodes. Note that the resulting sound pressure depends nonlinearly on the input signal due to dependence of $\mu$ on the strength of the field $E$. The
mobility $\mu$ itself is determined with respect to the static field, therefore, only the static constituent is taken into account in $\mu(E)$. The
calculations show (see Appendix) that for amplitudes of the alternating field $E/E_0<0.25$ the nonlinearities are small.

\section{The experimental setup}\setcounter{equation}{0}

\begin{figure}
  \begin{center}\includegraphics[scale=0.6]{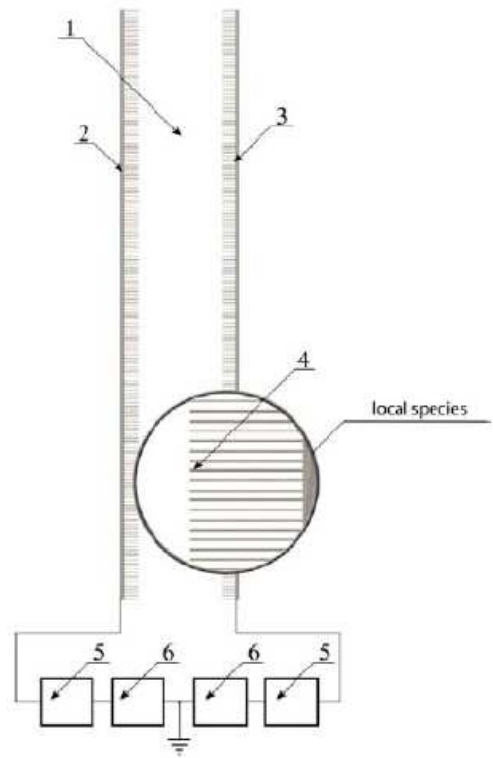}\\
  \caption{Scheme of the device.}\label{fig_needles}\end{center}
\end{figure}

\begin{figure}
  \begin{center}\includegraphics[scale=0.58]{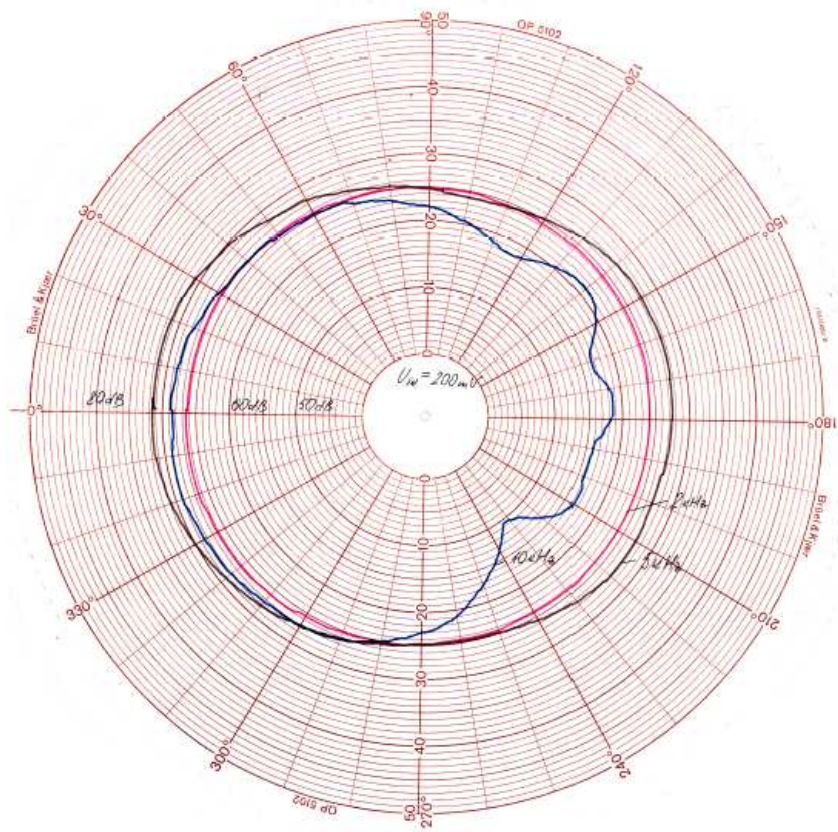}\\
  \caption{Directional pattern.}\label{fig_dirpattern}\end{center}
\end{figure}

\begin{figure}
  \begin{center}\includegraphics[scale=0.51]{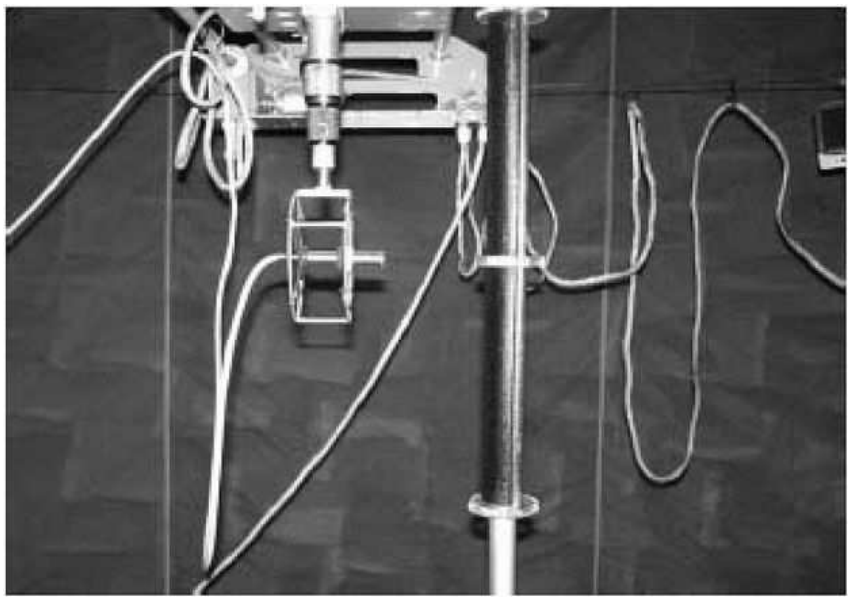}\\
  \caption{The measurement setup of the directivity pattern.}\label{fig_echo}\end{center}
\end{figure}

The ionic acoustic system (see Fig.~\ref{fig_needles}) consists of two arrays of thin needles (4) of stainless steel of the diameter 0,025 mm with the sequence
period compared with their diameter attached to two electrodes (2,3) of length 200 mm. Constant voltage (10-30 kV) under which the corona discharge exists is
applied on the electrodes. While modulating the value of this voltage within the limits of steady existence of the corona discharge, the generation of the
corresponding acoustic signal arises in the interelectrode interval (1). Electric measurements of VAC were carried out with the help of the oscillograph LeCroy
WR-62Xi. The constructed theoretical model describes the really created models of the so-called ionic acoustic systems. While working, the improved design,
representing two electrodes (2,3) in the form of arrays of thin wires of stainless steel of the diameter 0,025 mm (4), with the sequence period compared with
their diameter, was elaborated. The series constant-voltage sources (6) and grids (5) were used as sources.
On Fig.~\ref{fig_dirpattern} we depict the directional pattern of the device which shows that the first term really gives the main contribution at least for
the frequencies less than $10$ kHz. Measurements were carried out in the anechoic box with the help of the gauged electrical-type acoustic equipment
of ``Bruel \& Kjaer`` (see Fig.~\ref{fig_echo}).
The ionic acoustic system was set on the rotary basis, which slowly continuously rotated. Simultaneously the value of the amplitude of the generated sound was
registered for the frequencies 2, 5, 10 kHz. The microphone was located on the acoustic axis of the loud-speaker assemblage, at the distance 1 m.  The
directional pattern has some distortions for the frequency 10 kHz, which are explained by commensurability of the wave length and linear dimensions of the
loud-speaker elements, including the inner grid and the dielectric holders.

\section{Characteristic values and estimates of main quantities}\label{sec_estimates}\setcounter{equation}{0}

In this Section we derive some general inferences from the basic equation Eq.~(\ref{eq_sound_main}) and give the estimates for the relevant physical parameters
of the setup. It should be mentioned beforehand that the characteristic space scales are as follows: $L=2.1\, cm$ is the distance between electrodes and
$S=4\times10^{-3}\, cm^2$ is the typical value of the cathode surface area.

In order to get some estimates we neglect the coordinate dependence of the mobility $\mu$ and therefore the third term is omitted. This is based on the
following reasoning. The spatial variation of $\mu$ is caused by its dependence on the strength of the electric field $E$. The main contribution to the
mobility $\mu$ is due to the electronic component with the mobility $\mu_{e}$. For the characteristic values $E\simeq 10^4 \,\, V/cm$, $P_0= 1 \,\,atm$, under
which the device is operating, it does not depend essentially on $E$ according to \cite{book_reiserfizikagazrazrada_en}. This justifies the fact that the
spatial variation of $\mu$ can be neglected. The second term can be also neglected provided that $j_0$ is small, i.e. no substantial ion wind produced. Our
experimental setup confirms the possibility of such simplification.

Bearing in mind these simplifications and using the Poisson equation  \eqr{eq_electric} (we put $\varepsilon=1$ for simplicity), from \eqr{eq_sound_main}
we get another equivalent form:
\begin{equation}\label{eq_sound1}
\frac{1}{c^2}\frac{\partial^2P}{\partial t^2}-\triangle P=\frac{1}{4\pi \mu}\nabla\frac{\partial{\bf E}}{\partial t}\,.
\end{equation}
This form makes clear another direct inference that the phase shift between the excess sound pressure and the voltage of the harmonic signal is equal to $-\pi/2$.

The sound intensity can be estimated from the dimension reasoning. Indeed, the excess pressure produced by the bulk charge oscillation in the interval of
length $L$ between electrodes, reads as follows:
\begin{equation}\label{estim_pressure}
P_s\sim L\rho E_1\sim E_0E_1\, ,
\end{equation}
where $E_1$ is the amplitude of the signal. For the typical values $E_0\sim10\, un.\, CGS$, $E_1\sim0.1E_0$ we obtain $P_s\sim10\, un.\, CGS=1\, Pa$,
resulting in the sound pressure level
\begin{equation}\label{estim_soundintensity} I_s=20\, \mathrm{lg}\left(\frac{P_s}{20\, \mu
Pa}\right)\sim90\, dB\, .
\end{equation}
This simple estimate shows that in the considered electroacoustic system the sound pressure level can be made big enough. Below we refine the estimate basing
of the complete set of equations.

The obtained equations allow to estimate the value of  $\frac{\partial {\rho}}{\partial t}$ directly from \eqr{eq_kinetic}. Using the data of
\cite{book_plasmaprobojdjakov_en}, for the characteristic strength of the electric field we get $\nu^{(i)}_{+,e}\sim 10^5\,s^{-1},\,\,n_{e}\sim 10^5\,cm^{-3}$ and
therefore
\begin{equation}\label{estim_kinetic}
\frac{\partial {\rho}}{\partial t}\sim 5\, un.\, CGS\, .
\end{equation}
This leads to the value of the excess pressure $P_s \simeq 0.4 \,Pa$ and the corresponding sound intensity $I_s =86\,Db$. Comparison of the estimates
\eqr{estim_pressure} and \eqr{estim_kinetic} shows the self-consistency of the approach. As a result, we have shown that the generation of sound is due to the
oscillation of the bulk charge density.

Let us also note that the presence of attachment with formation of anions leads to the larger change of the charge density in comparison with the case of the
presence only of standard ionization/recombination with formation only of cations. Taking into account the large sensitivity of cross-sections of anion-aided
reactions to the field strength, one can draw a conclusion that the presence of electronegative ions promotes the power increase of electroacoustic generation
(see also \cite{plasma_soundamplif_zhetp1989_en}, where the mechanism of intensification of acoustic oscillations at the expense of friction between electrons
and neutral particles is proposed).

\section{Volt-ampere characteristic}\label{sec_device}\setcounter{equation}{0}

The primary generation of charge is induced by the static field with the near breakdown strength. In order to determine the mechanism of the primary injection
of the charges, we perform the measurements of the volt-ampere characteristic (VAC) for the stationary (no-operation) mode where the sound generation is
absent. The VAC of such mode is directly measurable.

\begin{figure}[hbt]
\begin{center}\includegraphics[scale=0.51]{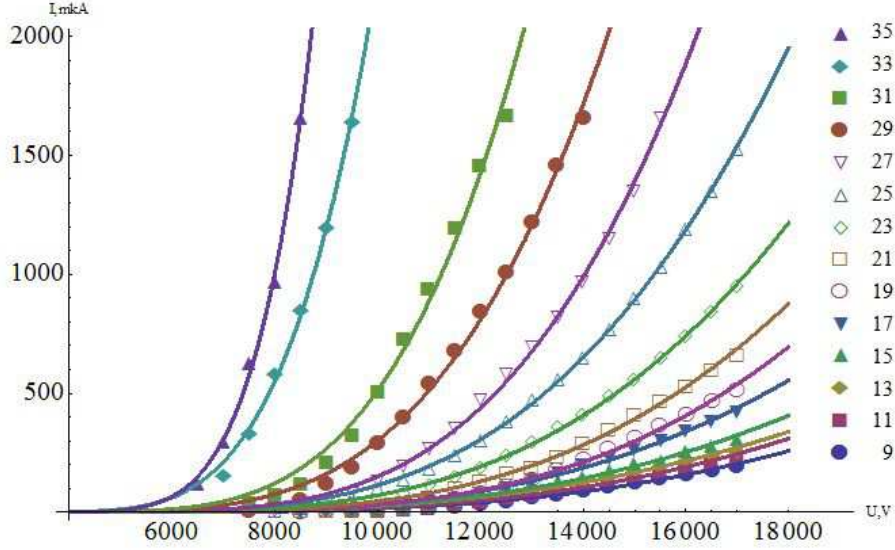}
\caption{Comparison of results of approximation of the experimental VAC by Eq.~(\ref{eq_autoelectremiss_fit}).}\label{fig_vacemiss}
\end{center}
\end{figure}

The results of the VAC measuring correlate well with the VAC of autoelectronic emission (that is explicitly demonstrated on Fig.~\ref{fig_vacemiss}).

We use the standard expression for the autoelectronic emission current density (see, e.g., \cite{plasma_autoelemisselinson_en,plasma_booklathamhighvolt})
\begin{equation}\label{eq_autoelectremiss}
j = j_*\,(E_0/E_*)^2\exp\left(-\frac{E_*}{E_0}\right)\, .
\end{equation}
In reality $j_*$ and $E_*$ differ very much from the theoretical values because of influence of the surface roughening on the work function
\cite{plasma_booklathamhighvolt}. In general these parameters strongly depend on the geometry of the electrodes.  Integrating \eqr{eq_autoelectremiss}
over the surface of the electrode, we get the VAC for the total current:
\begin{equation}\label{eq_autoelectremiss_fit}
I(U)=I_*\left(\frac{U}{U_*}\right)^2\exp\left(-\frac{U_*}{U}\right)\,,
\end{equation}
where $I_*$ and $U_*$ are some fitting parameters which depend on the distance $d$ between the electrodes. It should be noted that the demonstrated correlation
between the experimental VAC and the theoretical autoelectronic emission VAC \eqr{eq_autoelectremiss_fit} allows to use it as a boundary condition in the
solution of the problem (which is discussed in detail by the example of the one-dimensional model in Appendix).

\begin{figure}
\begin{center}\subfigure[\ The parameter $I_*(d)$]{\includegraphics[scale=0.625]{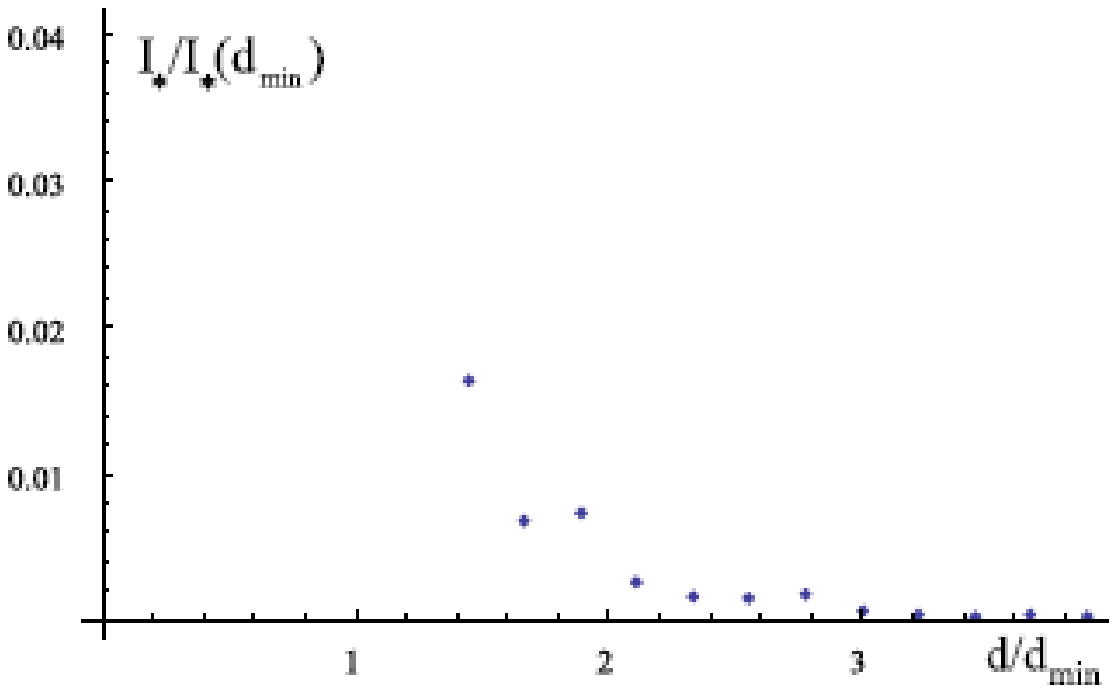}}\quad \subfigure[\ The parameter
$U_*(d)$]{\includegraphics[scale=0.625]{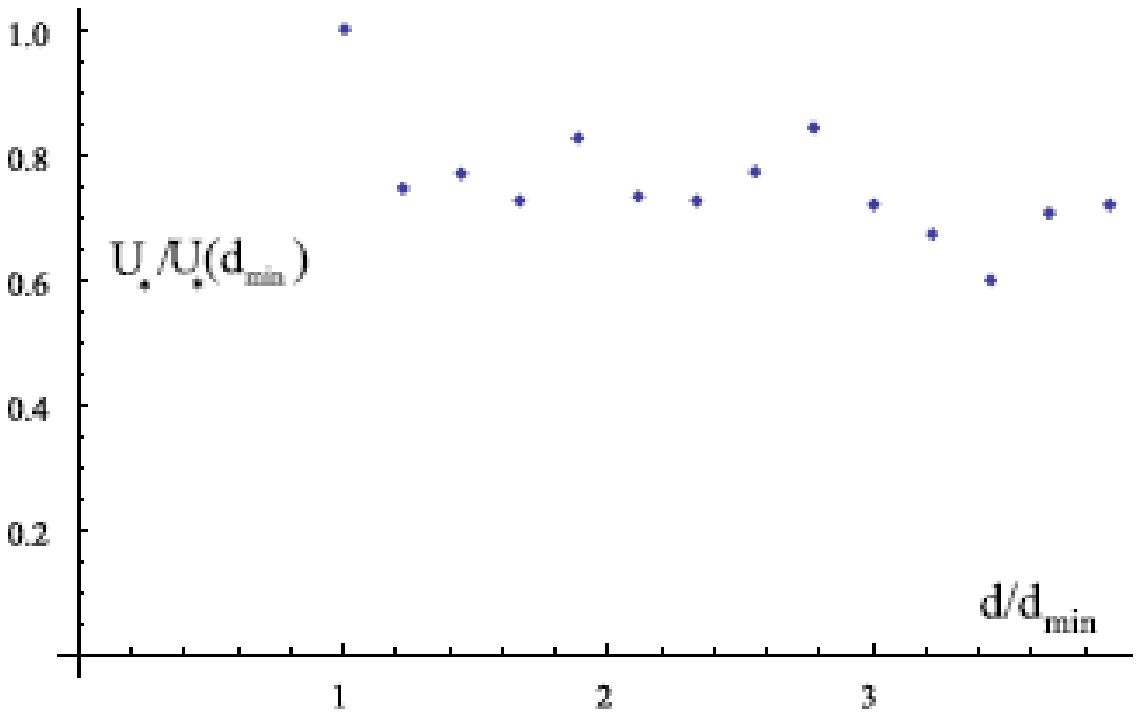}}
  \caption{Dependence of parameters $U_*$ and $I_*$ on $d$.}\label{fig_je}\end{center}
\end{figure}

\begin{figure}
  \begin{center}\includegraphics[scale=0.5]{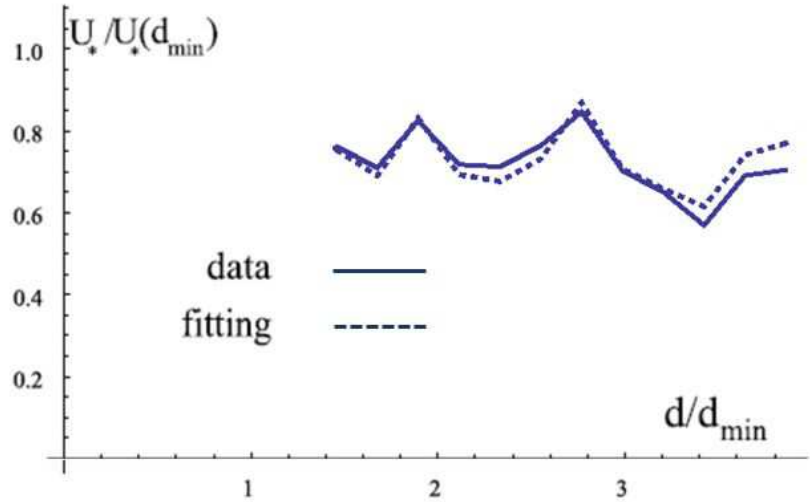}\\
  \caption{Comparison of $U_{*}$ and the approximation \eqr{jed}.}\label{fig_jed}\end{center}
\end{figure}

With the help of the experimental data on VAC we are able  to analyze the dependence of parameters $I_*, U_*$ on $d$. The results are shown on Fig.~\ref{fig_je}.
Since both fitting parameters $I_{*}$ and $U_{*}$ depend on distance $d$ there is the relation between them which is determined by the specific geometry of the
electrodes. With sufficient accuracy the following relation holds:
\begin{equation}\label{jed}
I_{*}(d)/I_{*}(d_{min}) \approx  0.33 \,\f{U_*(d)/U_*(d_{min}) -0.54 }{\left(\f{d}{d_{min}}\right)^4}\, .
\end{equation}
with $d_{min} = 9 \, mm$ and $d_{max} = 35 \,mm$. It is shown on Fig.~\ref{fig_jed}.
The dependence $I_{*}(d) \sim d^{-4}$ can be expected from the following reasoning.
It is well known that for the needle-plate geometry the current depends on the distance as $I\sim d^{-2}$ \cite{book_reiserfizikagazrazrada_en}.
Obviously, when the roughening is essential for both electrodes, the current depends on $d$ as $I\sim d^{-4}$, which is in agreement with \eqref{jed}. %

\section*{Conclusion}
In this paper we have described the sound wave generation in the gap between electrodes in the presence of the alternating electric field constituent. The
consideration is based on the simple three-component hydrodynamic model of the weakly ionized gas. It has been shown that for the cold plasma sound emitters
there is the oscillation of the bulk charge monopolar source of the sound. We have demonstrated that the bulk charge serves as the sound source and constructed
the theoretical scheme of calculation of the sound and its main characteristic: the excess sound pressure. To test the approach we give the solution for the 1D
model. The theoretical model proposed in the paper is aimed at the description of ``membrane-free`` acoustic systems which have no essential disadvantages
inherent to standard electrodynamic loud-speakers.

\appendix\section{One-dimensional model}

The case of the linear geometry is the simplest one, since all quantities depend only on one spatial coordinate (for example, $x$). Here we consider 1D analog
of the theoretical model described above and calculate the corresponding acoustic pressure. Despite the quite artificial restriction of planar 1D geometry it
could give useful results as shown in \cite{plasma_acoustdischwirewire_compmodel2009} where the idea of effective thickness and the corresponding reduction of
2D problem to 1D one has been proposed. Yet we do not resort to the effective thickness approach since the detailed theoretical model is out of the scope of
the present paper.

In the 1D case from \eqr{eq_electric} and  Eqs.~\eqref{eq_kinetic}, \eqref{eq_sound1} we obtain:
\begin{equation}\label{eq_poiss1dim}
\frac{\partial E_x}{\partial x}=4\pi e(n_{+}-n_{-}-n_e)\, ,
\end{equation}
\begin{equation}\label{eq_sound1dim}
 \frac{1}{c^2}\frac{\partial^2P}{\partial t^2}-\frac{\partial^2 P}{\partial x^2}=\frac{e}{\mu}\frac{\partial}{\partial t}(n_{+}-n_{-}-n_e)\,,
\end{equation}
and
\begin{eqnarray}\label{eq_kinetic1}
\frac{\partial n_{+}}{\partial t}+\frac{\partial}{\partial x}(n_{+}\mu_{+}E_x)&=&\nu_{+,e}^{(i)}n_e-\nu_{+,e}^{(r)}n_{+}n_{e}\, ,\nn\\
\frac{\partial n_{-}}{\partial t}-\frac{\partial}{\partial x}(n_{-}\mu_{-}E_x)&=&\nu_{0,e}^{(a)}n_e\, ,\nn\\
\frac{\partial n_{e}}{\partial t}-\frac{\partial}{\partial
x}(n_{e}\mu_{e}E_x)&=&\left(\nu_{+,e}^{(i)}-\nu_{0,e}^{(a)}\right)n_e-\nu_{+,e}^{(r)}n_{+}n_{e}\, ,\end{eqnarray}
respectively. If the distance between the electrodes is equal to $L=2.1\, cm$ and the voltage amounts to $\varphi_0=20\times 10^3/300\, un.\, CGS$ (or $20\,
kV$), then the characteristic value of the strength of the electric field amounts to approximately $32\, un.\, CGS$ (or $9.5\, kV/cm$). According to
\cite{book_reiserfizikagazrazrada_en}, $\mu_+=\mu_-=1.5\times10^{-4}\times(3\times10^6)=450\, un.\, CGS$,
\begin{equation}\label{mue} \mu_e\approx\left(0.027+\frac{0.113}{\sqrt{1+0.3E_0}}\right)\times(3\times10^6)\ un.\ CGS\, , \end{equation}
where $E_0$ denotes the absolute value of the continuous electric field strength being also expressed in CGS units, $\nu_{+,e}^{(r)}\approx
10^{-13}\times10^6=10^{-7}$,
\begin{eqnarray}\label{nu}
&{}&\nu_{+,e}^{(i)}\approx8\times10^{10}\times \exp\left(-\frac{13.2\times 10^{2}}{3E_0}\right)\ s^{-1},\quad \nonumber\\
&{}&\nu_{0,e}^{(a)}\approx 9.54\times10^9\times\exp\left(-\frac{7.32\times10^2}{3E_0}\right)\ s^{-1}\, . \end{eqnarray}
For the electric field strength characteristic value $32\, un.\, CGS$ we
get $\mu_e\approx1.85\times10^5$, $\nu_{+,e}^{(i)}\approx8.5\times10^4$ and $\nu_{0,e}^{(a)}\approx4.6\times10^6$.

\begin{figure}
\begin{center}\subfigure[\ $n_+(x)$]{\includegraphics[scale=0.62]{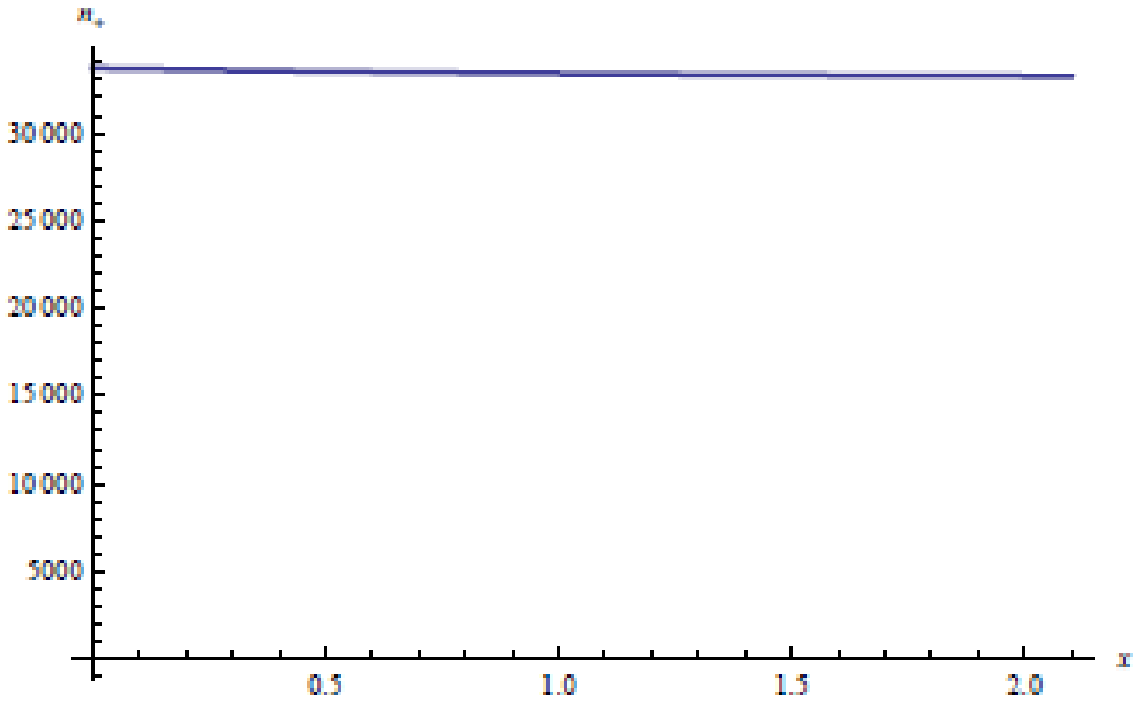}}\quad \subfigure[\ $n_-(x)$]{\includegraphics[scale=0.62]{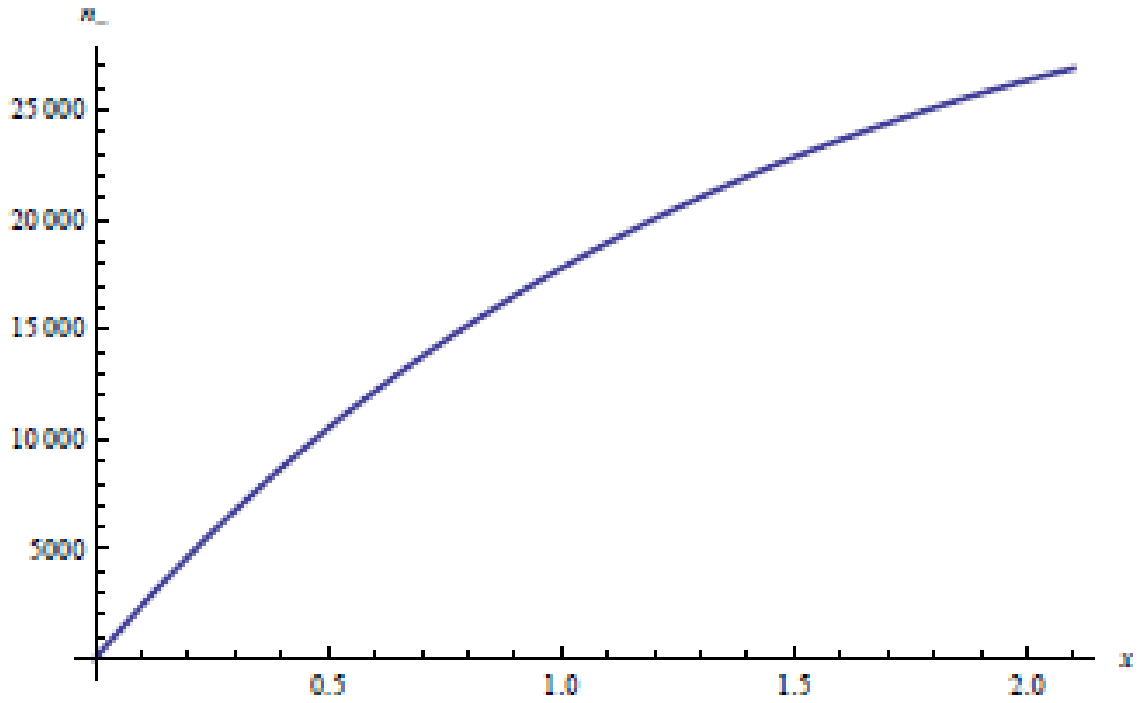}}
  \caption{Densities of positive and negative ions in stationary mode (see text).}\label{nplus_nminus}\end{center}
\end{figure}

\begin{figure}
\begin{center}\subfigure[\ $n_e(x)$]{\includegraphics[scale=0.61]{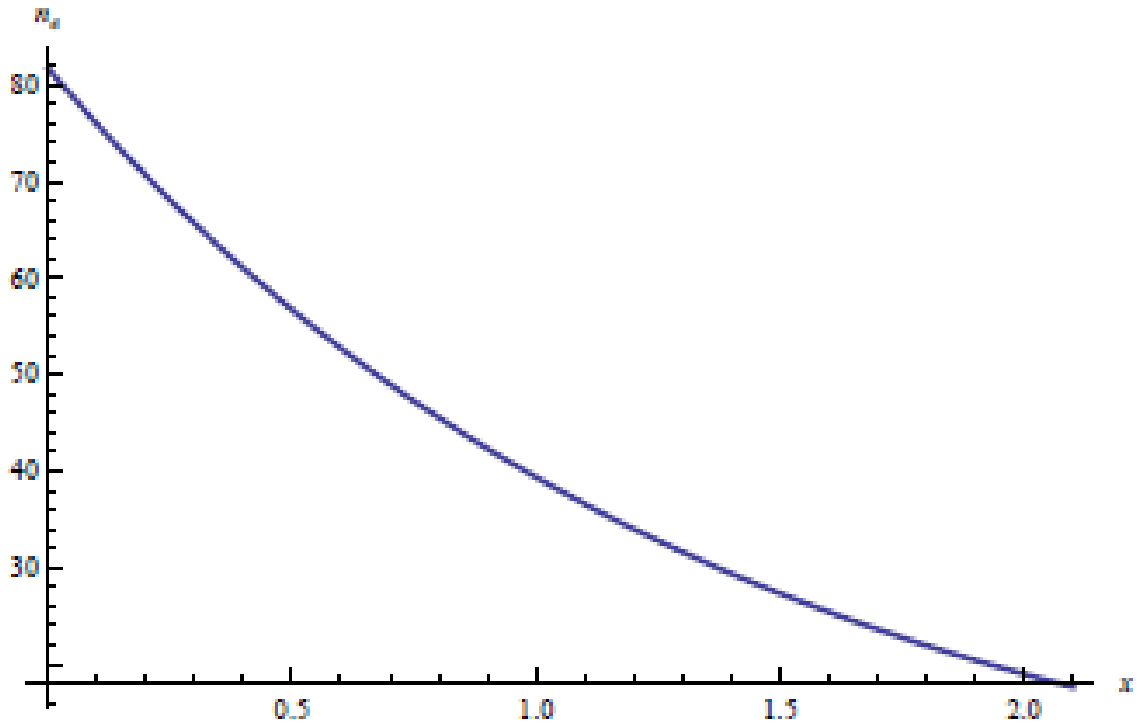}}\quad \subfigure[\ $\rho(x)$]{\includegraphics[scale=0.61]{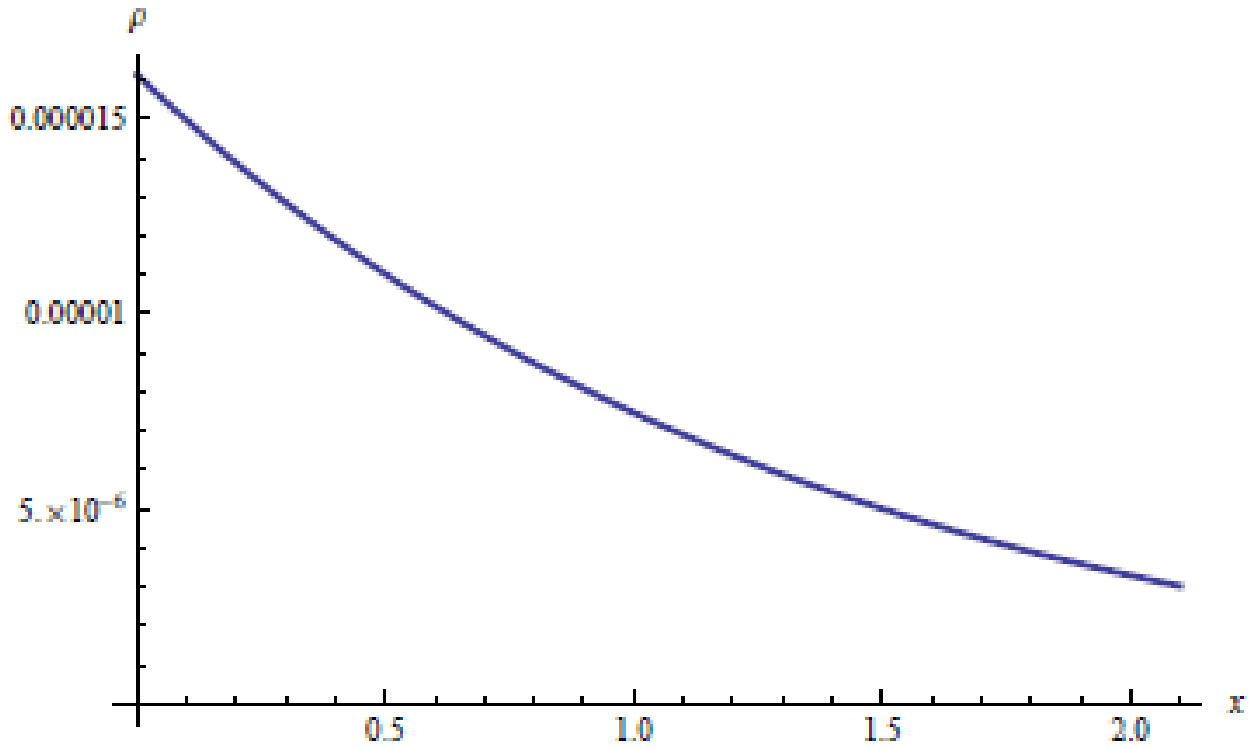}}
  \caption{Densities of electrons (a) and total charge (b) in stationary mode (see text).}\label{ne_rho}\end{center}
\end{figure}

Using the results of VAC measurements, for the value $S=4\times10^{-3}\, cm^2$ of the cathode surface area we solve Eqs. \eqref{eq_poiss1dim} and
\eqref{eq_kinetic1} numerically for stationary mode. The spatial distributions of species and the total charge are in Figs.~\ref{nplus_nminus},\ref{ne_rho}.

\begin{figure}
\begin{center}\subfigure[\ $n_+(x,t)$]{\includegraphics[scale=0.3]{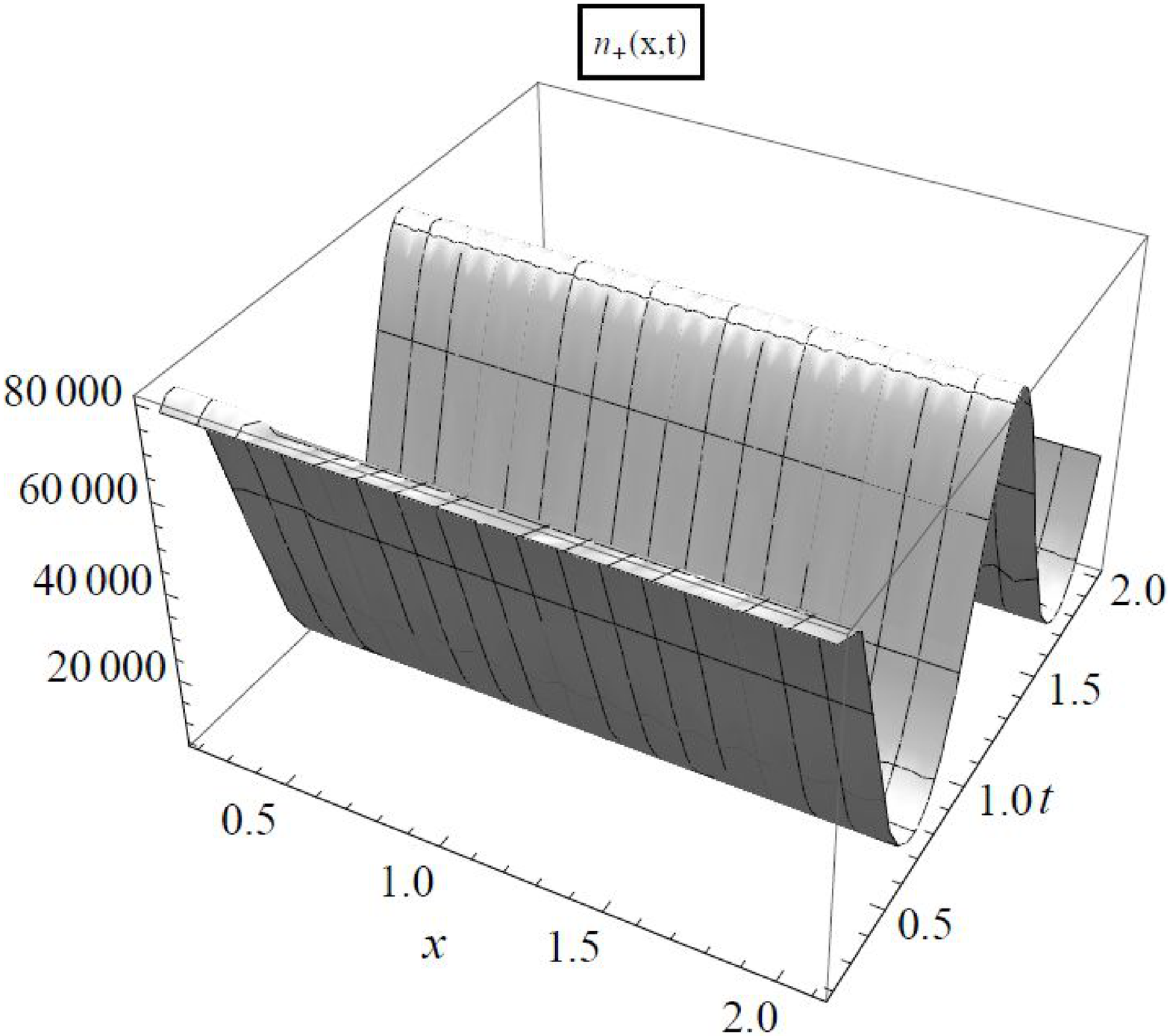}}\quad \subfigure[\ $n_-(x,t)$]{\includegraphics[scale=0.3]
{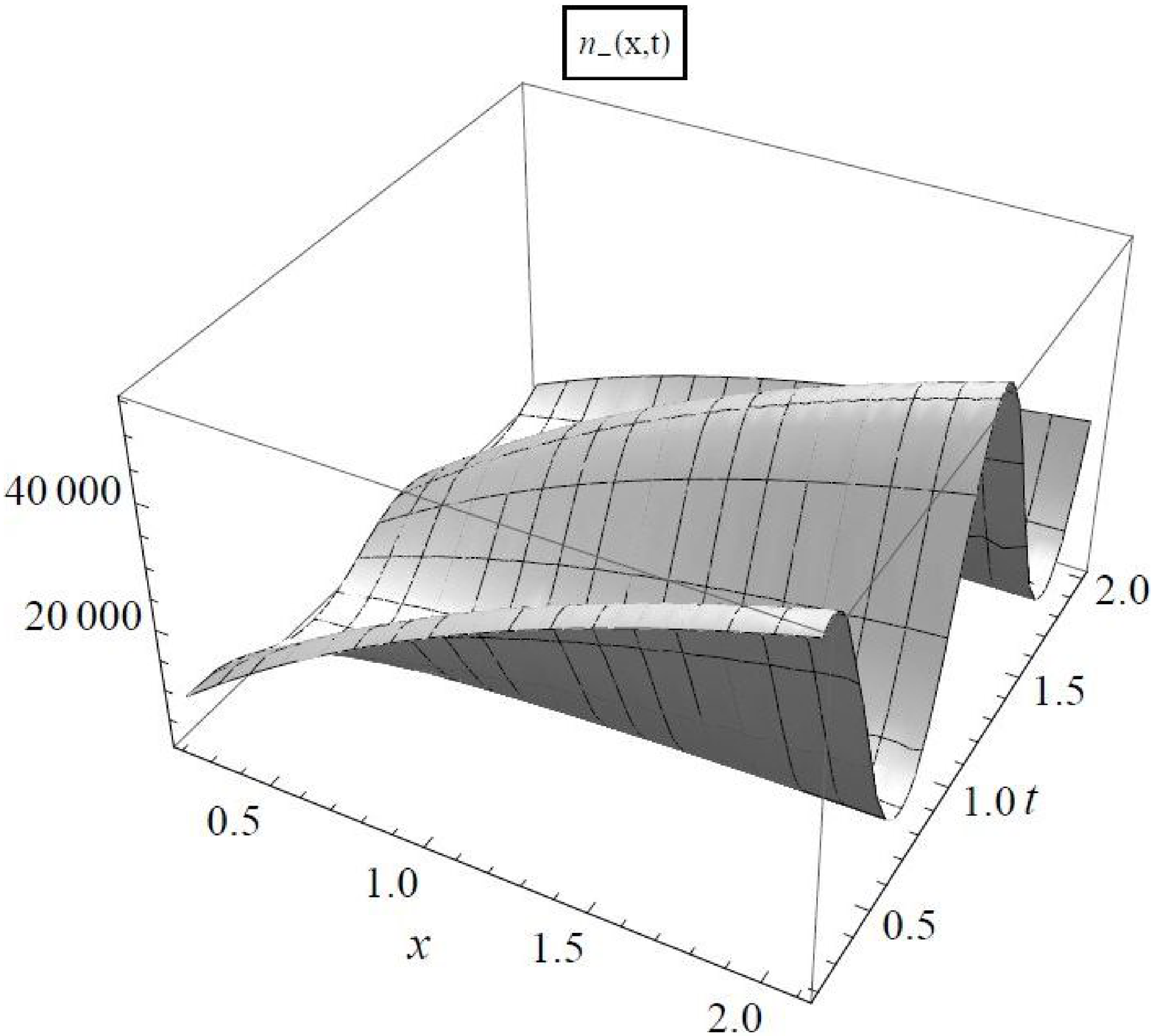}}
\caption{Densities of positive (a) and negative (b) ions in non-stationary mode \eqref{volt1}.}\label{fig_1dim3dcharges1}\end{center}
\end{figure}

\begin{figure}
\begin{center}\subfigure[\ $n_e(x,t)$]{\includegraphics[scale=0.3]{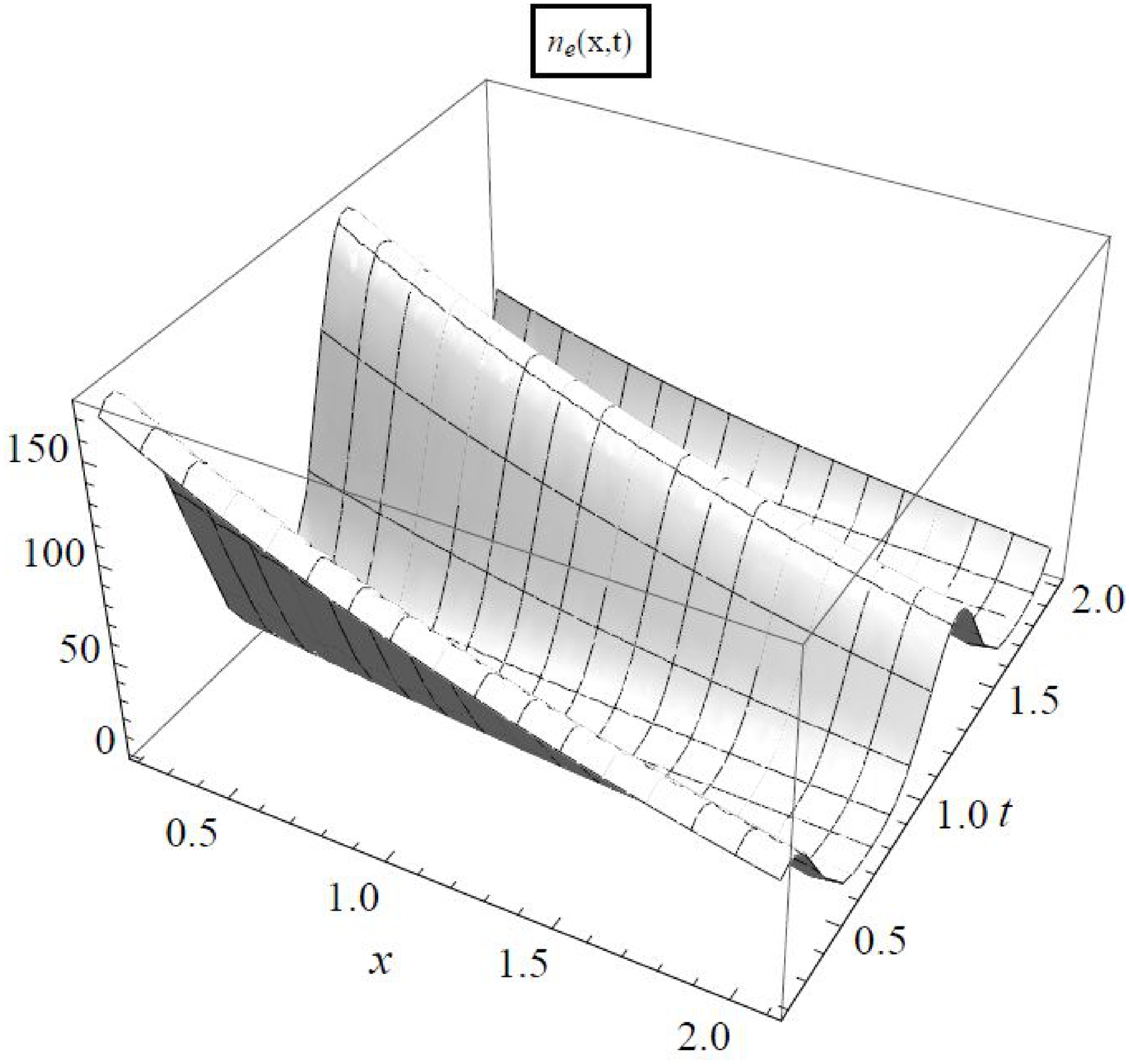}}\quad \subfigure[\
$\rho(x,t)$]{\includegraphics[scale=0.3]{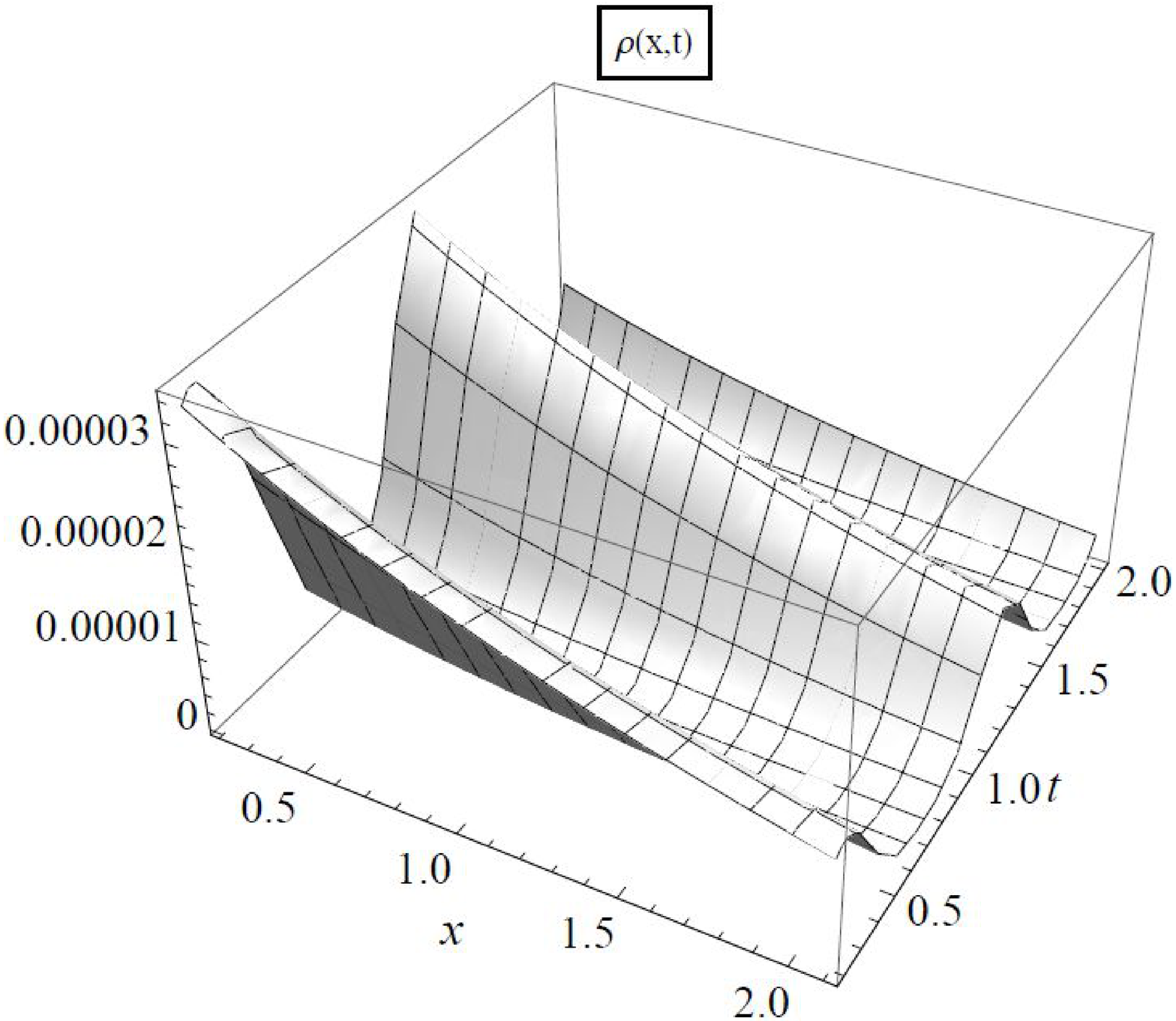}} \caption{Densities of electrons (a) and total charge (b) in non-stationary mode
\eqref{volt1}.}\label{fig_1dim3dcharges2}\end{center}
\end{figure}

The non-stationary mode is realized by applying the voltage of the form
\begin{equation}\label{volt1}
 \varphi=\varphi_0\,(1 +0.4\sin(\omega t))\,,
\end{equation}
where $\omega$ is the signal frequency, which for illustrative purposes is assumed to equal $2\pi\, s^{-1}$ (so, the period $T=2\pi/\omega=1\, s$). We solve
Eqs.~\eqref{eq_poiss1dim}-\eqref{eq_kinetic1} on the time interval $[0,2T]$. The 3D distributions of charges in $(x,t)$ variables are shown in
Fig.\ref{fig_1dim3dcharges1},\ref{fig_1dim3dcharges2}. According to the results, the corresponding sound intensity is
\begin{equation}\label{press_1d}
I_s = 20 \lg\left(\f{p}{p_{min}}\right)
\end{equation}
where the pressure can be estimated as
\begin{equation}\label{pressureforce}
 p = \int\limits_{0}^{L}\,\rho(x,t)\,E_{s}(x,t)\,dx\,.
\end{equation}
The calculation gives the result $I_s \approx 25$\,\, dB.

\newpage
\section*{References}
%
\bibliography{JPD_acoust_paper_March_2013_3.bbl}
\end{document}